\documentclass[12pt]{article}
\usepackage[dvips]{epsfig}

\begin{document}

\author{A. de Souza Dutra$^{a,b}$\thanks{%
E-mail: dutra@feg.unesp.br} \\
$^{a}$Abdus Salam ICTP, Strada Costiera 11, 34014 Trieste Italy\\
$^{b}$UNESP-Campus de Guaratinguet\'{a}-DFQ\thanks{%
Permanent Institution}\\
Av. Dr. Ariberto Pereira da Cunha, 333\\
C.P. 205\\
12516-410 Guaratinguet\'{a} SP Brasil}
\title{{\LARGE Ordering ambiguity versus representation}}
\maketitle

\begin{abstract}
In this work we show that the ordering ambiguity on quantization depends on
the representation choice. This property is then used to solve unambiguously
a particular system. Finally we speculate on the consequences for more
involved cases.

\end{abstract}

\newpage

The problem of ordering ambiguity is one of the long standing
questions of the quantum mechanics. This question has attracted
the attention of some of the founders of the quantum mechanics.
Born and Jordan, Weyl, Dirac and von Newmann worked on this
matter, as can be verified from the excellent review by Shewell
\cite{shewell}. This is a hard problem which has advanced very few
along the last decades. Notwithstanding, as a consequence of its
importance in some experimental situations like impurities in
crystals \cite {luttinger}-\cite{slater}, the dependence of
nuclear forces on the relative velocity of the two nucleons
\cite{rojo}, \cite{razavy}, and more recently the study of
semiconductor heterostructures \cite{bastard},\cite{weisbuch}, the
interest in such kind of systems never vanished. Furthermore,
taking in account the spatial variation of the semiconductor type,
some effective Hamiltonians were proposed with a spatially
dependent mass for the carrier \cite{bendaniel}-\cite{cavalcante}.
Some time ago we discussed the exact solvability of some classes
of Hamiltonians with ordering ambiguity \cite {dutra1}. In fact,
the problem of the spatially dependent mass is presenting a
growing interest along the last few years
\cite{dutra1}-\cite{epl05}.

Let us present the idea we are interested in develop in this work by
illustrating it through the study a concrete example. In \cite{dutra1} it
was shown that for a system whose quantum Hamiltonian has as one of its
parts an operator version of the classical function $f\left( x\right) \,p$.
In the coordinate representation its operator counterpart can be written as
\begin{equation}
f\left( x\right) \,p\,\rightarrow \frac{f^{\alpha }\left( \hat{x}\right) \,%
\hat{p}\,f^{\beta }\left( \hat{x}\right) +f^{\beta }\left( \hat{x}\right) \,%
\hat{p}\,f^{\alpha }\left( \hat{x}\right) }{2},  \label{e1}
\end{equation}

\noindent where $\alpha +\beta \equiv 1$. By using the usual coordinate
representation for the operator $\hat{p}$, and manipulating the above
operator in order to put it at right, one can see that one obtains for
instance
\begin{equation}
f^{\alpha }\left( \hat{x}\right) \,\hat{p}\,f^{\beta }\left( \hat{x}\right)
=f\left( \hat{x}\right) \,\hat{p}\,-i\,\hbar \,\beta \,\frac{df\left( \hat{x}%
\right) }{d\hat{x}}\,.
\end{equation}

\noindent Now, using the corresponding operator for $f^{\beta }\left( \hat{x}%
\right) \,\hat{p}\,f^{\alpha }\left( \hat{x}\right) $, and then calculating
the Hermitian operator (\ref{e1}) with these features, one gets finally
\begin{equation}
\frac{f^{\alpha }\left( \hat{x}\right) \,\hat{p}\,f^{\beta }\left( \hat{x}%
\right) +f^{\beta }\left( \hat{x}\right) \,\hat{p}\,f^{\alpha }\left( \hat{x}%
\right) }{2}=f\left( \hat{x}\right) \,\hat{p}\,-\frac{i\,\hbar }{2}\,\,\frac{%
df\left( \hat{x}\right) }{d\hat{x}},\,
\end{equation}

\noindent from which we conclude that there is no ordering ambiguity in this
representation and any ordering used will conduce essentially to the same
final answer, as observed in \cite{dutra1}. However, despite of being an
important case of ordering, due to its application in the case of the
minimal gauge coupling, it can not be used itself as a Hamiltonian, at least
an usual one, because the momentum appears linearly in it. At this point we
introduce the main idea underlying this work, by remembering that one could
interchanges the role of $x$ and $p$, and discussing the case of the
quantization of the classical function $g\left( p\right) \,x$ in the
momentum representation. It is not hard to conclude, through an absolutely
analoguous analysis in the momentum representation that the Hermitian
quantization of this function is unambiguous, and looks like
\begin{equation}
\frac{g^{\alpha }\left( \hat{p}\right) \,\hat{x}\,\,f^{\beta }\left( \hat{p}%
\right) +f^{\beta }\left( \hat{p}\right) \,\hat{x}\,\,f^{\alpha }\left( \hat{%
p}\right) }{2}=f\left( \hat{p}\right) \,\hat{x}\,+\frac{i\,\hbar }{2}\,\,%
\frac{df\left( \hat{p}\right) }{d\hat{p}}.
\end{equation}

\noindent Note, however, that this operator is surely ambiguous in the
coordinate representation. From the above calculation we can conclude that
the ordering ambiguity has a dependence on the choice of representation and,
as far we know, this feature was not taken into account in the literature up
to now. Furthermore, this last operator can be thought as a Hamiltonian if
we choose $g\left( \hat{p}\right) =\hat{p}^{2}$. In this special case, we
would have a system with spatial dependence in the spatial coordinate ($%
m\left( x\right) \sim \frac{1}{x}$). This is an example of Hamiltonian which
is ambiguous in the coordinate representation and not in the momentum one.
In cases like this, one could calculate the wave function in the momentum
representation and then transform it trough
\begin{equation}
\psi \left( x,t\right) =\int dp\,\tilde{\psi}\left( p,t\right) \,e^{\frac{i}{%
\hbar }\,p\,x},
\end{equation}

\noindent to the coordinate representation if necessary.

For the sake of concreteness, from now on we discuss this case with more
details. Firstly, the time-independent wave function equation in the
momentum representation is given by
\begin{equation}
i\,\hbar \,p^{2}\,\frac{d\tilde{\psi}\left( p\right) }{dp}+i\,\hbar \,p\,%
\tilde{\psi}\left( p\right) =E\,\mathit{\,}\tilde{\psi}\left( p\right) .
\end{equation}

After a straightforward calculation, one obtains for the unambiguous wave
function in the momentum representation
\begin{equation}
\tilde{\psi}\left( p\right) =N\,\frac{\,e^{\left( \frac{i}{\hbar }\frac{E}{p}%
\right) }}{p},
\end{equation}

\noindent where $N$ is an arbitrary integration constant. We can now
calculate its Fourier transform, in order to obtain the corresponding
coordinate representation wave function. So, we must perform the following
integration
\begin{equation}
\psi \left( x\right) =N\,\,\int_{-\infty }^{\infty }\frac{dp}{p}%
\,\,e^{\left( \frac{i}{\hbar }\frac{E}{p}+p\,x\right) }.
\end{equation}

\noindent In order to reach this goal, we separate the integral in two
sectors, that for positive $p$ and that for negative ones. So we gets
\begin{equation}
\psi \left( x\right) =N\,\,\left\{ \int_{-\infty }^{0}\frac{dp}{p}%
\,\,e^{-\left( \frac{i}{\hbar }\frac{E}{p}+p\,x\right) }+\int_{0}^{\infty }%
\frac{dp}{p}\,\,e^{\left( \frac{i}{\hbar }\frac{E}{p}+p\,x\right) }\right\} ,
\end{equation}
which after some manipulations can be rewritten as
\begin{equation}
\psi \left( x\right) =2\,i\,N\,\,\int_{0}^{\infty }\frac{dp}{p}\,\,\sin
\left( \frac{i}{\hbar }\frac{E}{p}+p\,x\right) .
\end{equation}
Then, after using usual trigonometric identities and the known result
\begin{equation}
\int_{0}^{\infty }\frac{du}{u}\,\sin \left( a\,u\right) \,\cos \left( \frac{b%
}{u}\right) =\frac{\pi }{2}\,J_{0}\left( 2\,\left( a^{2}b^{2}\right) ^{\frac{%
1}{4}}\right) =\int_{0}^{\infty }\frac{du}{u}\,\sin \left( \frac{b}{u}%
\right) \,\cos \left( a\,u\right) ,
\end{equation}
where $J_{0}\left( z\right) $ is the Bessel function of first kind. One
obtains finally that
\begin{equation}
\psi \left( x\right) =2\,\pi \,i\,N\,\,J_{0}\left( \frac{2}{\hbar }\,\sqrt{%
\left| E\,x\right| }\right) .
\end{equation}

Have we started in the coordinate representation, the wave function equation
to be solved would be
\begin{equation}
x^{2}\frac{d^{2}\psi \left( x\right) }{dx^{2}}+x\,\frac{d\psi \left(
x\right) }{dx}-\alpha \,\gamma \,\psi \left( x\right) =-\left( \frac{E}{%
\hbar ^{2}}\right) \,x\,\psi \left( x\right) ,
\end{equation}

\noindent where we ordered the operator coming from $x\,p^{2}$ using
\begin{equation}
\mathit{O}_{p}\equiv \frac{1}{2}\left( \hat{x}^{\alpha }\,\hat{p}\,\hat{x}%
^{\beta }\,\hat{p}\,\hat{x}^{\gamma }+\hat{x}^{\gamma }\,\hat{p}\,\hat{x}%
^{\beta }\,\hat{p}\,\hat{x}^{\alpha }\,\right) =\hat{x}\,\hat{p}%
^{2}-i\,\hbar \,\,\hat{p}\,+\alpha \,\gamma \,\hat{x}^{-1},
\end{equation}

\noindent and we used $\alpha +\beta +\gamma =1$. It can be noted that, if
make the variable transformation $\left| x\right| =\frac{\hbar ^{2}}{4\,E}%
\,w $, the above equation can be cast in the form
\begin{equation}
w^{2}\frac{d^{2}\psi }{dw^{2}}+w\,\frac{d\psi }{dw}+\left( w^{w}-4\,\alpha
\,\gamma \right) \psi =0,
\end{equation}

\noindent which is the differential equation of the first kind Bessel
function. So, we get finally that, in the coordinate representation the
ambiguous wave function is expressed as
\begin{equation}
\psi \left( x\right) =\,N\,J_{\alpha \,\gamma }\,\left( \frac{2}{\hbar }%
\sqrt{\left| E\,x\right| }\right) ,
\end{equation}

\noindent once $E$ is positive definite. We conclude that the compatibility
of the solutions coming from the two representations, imposes us to fix one
of the parameters appearing in the index of the Bessel function ( $\alpha =0$
or $\gamma =0$) in the coordinate representation. As a consequence of the
symmetry between these parameters in the operator definition, is equivalent
to choose any of them equal to zero. Choosing to make $\gamma =0$, we
conclude that $\beta =1-\gamma $, and we ends with a subclass of operators,
compatibles in both representations,
\begin{equation}
\mathit{O}_{\alpha }=\frac{1}{2}\left( \hat{x}^{\alpha }\,\hat{p}\,\hat{x}%
^{1-\alpha }\,\hat{p}+\,\hat{p}\,\hat{x}^{1-\alpha }\,\hat{p}\,\hat{x}%
^{\alpha }\,\right) .
\end{equation}

\noindent Note that the case of the Li and Khun ordering \cite{li}, which we
have shown to be equivalent to the Weyl ordering \cite{dutra1}, corresponds
to the choice $\alpha =\frac{1}{2}$.

Below we are going to prove that, in fact, there is no remaining ambiguity
because all choices of $\alpha $ are equivalent. For this we note that
\begin{eqnarray}
\hat{x}^{\alpha }\,\hat{p}\,\hat{x}^{1-\alpha }\,\hat{p} &=&\sqrt{\hat{x}}\,%
\hat{p}\,\sqrt{\hat{x}}\,\,\hat{p}+i\,\hbar \left( \alpha -\frac{1}{2}%
\right) \hat{p};  \nonumber \\
&& \\
\,\,\,\hat{p}\,\hat{x}^{1-\alpha }\,\hat{p}\,\hat{x}^{\alpha } &=&\,\hat{p}\,%
\sqrt{\hat{x}}\,\,\hat{p}\,\sqrt{\hat{x}}-i\,\hbar \left( \alpha -\frac{1}{2}%
\right) \hat{p},  \nonumber
\end{eqnarray}

\noindent so that the operator $\mathit{O}_{\alpha }$ is simply rewritten as
\begin{equation}
\mathit{O}_{\alpha }=\frac{1}{2}\left( \,\sqrt{\hat{x}}\,\hat{p}\,\sqrt{\hat{%
x}}\,\,\hat{p}+\,\hat{p}\,\sqrt{\hat{x}}\,\,\hat{p}\,\sqrt{\hat{x}}\right) =%
\mathit{O}_{Weyl}.
\end{equation}

So we have finally demonstrated that, at least for this particular
case, we have been able to avoid the ordering ambiguity by working
in the momentum representation. In fact, this conclusion is still
true if we include a binding potential energy in the original
Hamiltonian. One can show also that, for a given class of
potentials, the problem can be even exactly solvable. Furthermore
we have shown that this unambiguous quantization corresponds to
the so called Weyl ordering.

It is interesting to see that, in some very recent papers, it was adjudicate
in favor of a Schroedinger equation in a phase-space representation, where
appears a very interesting kind of mixing between the usual coordinate and
momentum representations \cite{phase1},\cite{phase2}. It would be very
interesting to see if  this generalized representation could be useful in
some particular problem, where there exist the with usual ordering ambiguity
both in coordinate and momentum representations and, maybe not in this new
representation.

\bigskip

\noindent \textbf{Acknowledgments:} The author is grateful to CNPq for
partial financial support and to the professor M. Hott for many helpful
discussions. This work has been done during a visit under the auspices of
the Associate Scheme of the Abdus Salam ICTP.


\newpage

\begin{thebibliography}{99}
\bibitem{shewell}  J. R. Shewell, Am. J. Phys. \textbf{27} (1959) 16.

\bibitem{luttinger}  J. M. Luttinger and W. Kohn, Phys. Rev. \textbf{97}
(1955) 869.

\bibitem{wannier}  G. H. Wannier, Phys. Rev. \textbf{52} (1957) 191.

\bibitem{slater}  J. C. Slater, Phys. Rev. \textbf{76} (1949) 1592.

\bibitem{rojo}  \'{O}. Rojo and J. S. Levinger, Phys. Rev. \textbf{123}
(1961) 2177.

\bibitem{razavy}  M. Razavy, G. Field and J. S. Levinger, Phys. Rev. \textbf{%
125} (1962) 269.

\bibitem{bastard}  G. Bastard, Wave Mechanics Applied to Semiconductor
Heterostructres, Les \'{E}ditions de Physique, Les Ullis, 1992.

\bibitem{weisbuch}  C. Weisbuch and B. Vinter, Quantum Semiconductor
Heterostructures, Academic Press, New York, 1993.

\bibitem{bendaniel}  D. J. BenDaniel and C. B. Duke, Phys. Rev. B \textbf{152%
} (1966) 683.

\bibitem{zhu}  Q. G. Zhu and H. Kroemer, Phys. Rev. B \textbf{27} (1983)
3519.

\bibitem{gora}  T. Gora and F. Williams, Phys. Rev. \textbf{177} (1969) 1179.

\bibitem{bastard2}  G. Bastard. Phys. Rev. B \textbf{24} (1981) 5693.

\bibitem{li}  T. L. Li and K. J. Kuhn, Phys. Rev. B \textbf{47} (1993) 12760.

\bibitem{cavalcante}  F. S. A. Cavalcante, R. N. Costa Filho, J. Ribeiro
Filho, C. A. S. de Almeida and V. N. Freire, Phys. Rev. B \textbf{55} (1997)
1326.

\bibitem{dutra1}  A. de Souza Dutra and C. A. S. de Almeida, Phys. Lett. A
\textbf{275} (2000) 25.

\bibitem{dutra2}  A. de Souza Dutra, M. B. Hott and C. A. S. Almeida,
Europhys. Lett. \textbf{62} (2003) 8.

\bibitem{ufce}  F. S. A. Cavalcante, R. N. Costa Filho, J. Ribeiro Filho, C.
A. S. de Almeida and V. N. Freire, Phys. Rev. B \textbf{55} (1997) 1326.

\bibitem{cg1}  G. Chen, Chin. Phys. \textbf{14} (2005) 460.

\bibitem{Cg2}  G. Chen and Z. D. Chen, Phys. Lett. A \textbf{331} (2004) 312.%
{\small \ }

\bibitem{a0}  A. A. Stahlhofen, J. Phys. A \textbf{37} (2004) 10129-10138,

\bibitem{a1}  B. Bagchi, P. Gorain, C. Quesne and R. Roychoudhury, Mod.
Phys. Lett. A \textbf{19} (2004) 2765-2775,

\bibitem{a3}  K. Bencheikh, K. Berkane and S. Bouizane, J. Phys. A \textbf{37%
} (2004) 10719-10725,

\bibitem{a4}  J. A. Yu, S. H. Dong, Phys. Lett. A \textbf{325} (2004)
194-198,

\bibitem{a5}  C, Quesne and V. M. Tkachuk, J. Phys. A \textbf{37} (2004)
4267-4281,

\bibitem{a6}  Y. C. Ou, Z. Q. Cao and Q. H. Shen, J. Phys. A \textbf{37}
(2004) 4283-4288,

\bibitem{a7}  R. Koc and H. Tutunculer, Annalen der Physik \textbf{12}
(2003) 684-691,

\bibitem{a8}  A. D. Alhaidari, Phys. Rev. A \textbf{66} (2002) 042116,

\bibitem{a9}  B. Roy and P. Roy, J. Phys. A \textbf{35} (2002) 3961-3969.

\bibitem{npb}  S. Ramgoolam, B. Spence and S. Thomas, Nucl. Phys. B \textbf{%
703} (2005) 236.

\bibitem{epl05} A. de Souza Dutra, M. B. Hott and V. G. C. S. dos
Santos, Europhys. Lett. {\bf 71} (2005) 166.

\bibitem{phase1}  Q. S. Li, G. M. Wei and L. Q. L\"{u}, Phys. Rev A \textbf{%
70} (2004) 022105.

\bibitem{phase2}  D. Chru\'{s}ci\'{n}ski and K. Mlodawski, Phys. Rev.
A \textbf{71} (2005) 052104.
\end{thebibliography}
\end{document}